\pdfoutput=1
\RequirePackage{ifpdf}
\ifpdf % We are running pdfTeX in pdf mode
\documentclass[pdftex]{sigma}
\else
\documentclass{sigma}
\fi

\begin{document}

\allowdisplaybreaks

\renewcommand{\thefootnote}{$\star$}

\renewcommand{\PaperNumber}{060}

\FirstPageHeading

\ShortArticleName{Ladder Operators for Quantum Systems Conf\/ined by Dihedral Angles}

\ArticleName{Ladder Operators for Quantum Systems\\ Conf\/ined by Dihedral Angles\footnote{This
paper is a contribution to the Special Issue ``Superintegrability, Exact Solvability, and Special Functions''. The full collection is available at \href{http://www.emis.de/journals/SIGMA/SESSF2012.html}{http://www.emis.de/journals/SIGMA/SESSF2012.html}}}

\Author{Eugenio LEY-KOO~$^\dag$ and Guo-Hua SUN~$^\ddag$}

\AuthorNameForHeading{E.~Ley-Koo and G.-H.~Sun}

\Address{$^\dag$~Instituto de F\'{\i}sica, Universidad Nacional Aut\'{o}noma de M\'{e}xico, M\'{e}xico}
\EmailD{\href{mailto:eleykoo@fisica.unam.mx}{eleykoo@fisica.unam.mx}}

\Address{$^\ddag$~Centro Universitario Valle de Chalco, Universidad Aut\'{o}noma del Estado de M\'{e}xico, M\'{e}xico}
\EmailD{\href{mailto:sunghdb@yahoo.com}{sunghdb@yahoo.com}}

\ArticleDates{Received June 29, 2012, in f\/inal form September 07, 2012; Published online September 12, 2012}

\Abstract{We report the identif\/ication and construction of raising and lowering
operators for the complete eigenfunctions of isotropic harmonic oscillators
conf\/ined by dihedral angles, in circular cylindrical and spherical
coordinates; as well as for the hydrogen atom in the same situation of
conf\/inement, in spherical, parabolic and prolate spheroidal coordinates.
The actions of such operators on any eigenfunction are examined in the
respective coordinates, illustrating the possibility of generating the
complete bases of eigenfunctions in the respective coordinates for both
physical systems. The relationships between the eigenfunctions in each
pair of coordinates, and with the same eigenenergies are also illustrated.}


\Classification{20C35; 33C47; 35J15; 35J25; 81Q05; 81R05}

\section{Introduction}

In this section, we provide the background information on our investigations
on the hydrogen atom in diverse situations of conf\/inement motivated by a~variety of physical and chemical si\-tua\-tions, spanning a period of
thirty-some years as reviewed in \cite{1}, and also our current work for
conf\/inement by a dihedral angle~\cite{2}. In each case the Schr\"{o}dinger
equation is separable in the chosen coordinate system, and its solutions
dif\/fer from the familiar ones for the free hydrogen atom by the condition of
vanishing at the position of the conf\/ining natural boundary, determined by
the f\/ixed value of one of the coordinates.  Such ``natural'' boundaries allow
for separable and integrable solutions, with non-integer eigenvalues for the
constant of motion canonically conjugate to the conf\/ining coordinate.  The
presence of such eigenvalues in the dif\/ferential equations for the other two
degrees of freedom af\/fects correspondingly their eigensolutions, which
nevertheless maintain their respective integer quantum numbers and constants
of motion.The following paragraphs contain a description of our motivation
to search, identify and construct ladder operators for both the harmonic
oscillator and the hydrogen atom conf\/ined by dihedral angles, in the
dif\/ferent coordinate systems sharing the broken rotational symmetry around
the edge of the angle.

The research reported in this article was motivated by the feedback on our
contribution in the Symposium on Superintegrability, Exact Solvability, and
Special Functions. In fact, that contribution with the title
``Superintegrability in Conf\/ined Quantum Systems'' was focused on the case of
the hydrogen atom conf\/ined inside spheres, and prolate spheroids, and in
semi-inf\/inite spaces with paraboloidal, hyperboloidal and circular cone
boundaries~\cite{1} as well as dihedral angles~\cite{2}. The successive boundary
condition equations determining the eigenvalues of the respective constants
of motion canonically conjugate to the conf\/ining coordinate are%
\begin{gather}
M(-\nu _{r},  2\ell +2, 2r_{0}/\nu ) =0, \qquad
 \nu =\nu _{r}+\ell +1,\label{eq1}
\\
S_{\nu _{u}}(u_{0}) =0, \qquad
 \nu  =\nu _{u}+n_{v}+m+1,\label{eq2}
\\
M\big({-}\nu _{\xi }, m +1,  \xi _{0}^{2}/\nu \big) =0,\qquad
 \nu  =\nu _{\xi }+n_{\eta }+m+1, \label{eq3}
\\
S_{\nu _{v}}(v_{0}) =0,\qquad
 \nu  =n_{u}+\nu _{v}+m+1,\label{eq4}
\\
P_{\lambda }^{m}(\cos \vartheta _{0}) =0,\qquad
 \nu  =n_{r}+\lambda +1,\label{eq5}
\\
\sin \mu \varphi _{0} =0, \qquad
 \nu  =n_{r}+n_{\theta }+\mu+1=n_{u}+n_{v}+\mu +1=n_{\xi }+n_{\eta }+\mu +1,\label{eq6}
\end{gather}
 involving conf\/luent hypergeometric functions $M(a,b,z)$, prolate
spheroidal functions $S$ in the spheroidal $1<u_{0}<\infty $ and
hyperboloidal $-1\leqslant v_{0}\leqslant 1$ coordinates, associated
Legendre~$P_{\lambda }^{m}$ functions, and trigonometric sine functions.
Their respective quantum labels $\nu _{r}$, $\nu _{u}$, $\nu _{\xi }$, $\nu
_{v}$, $\lambda $ and~$\mu$ are no longer integer. Consequently, the
functions themselves are no longer polynomials as in the free hydrogen atom,
and the principal quantum label~$\nu $ is no longer an integer, either. The
energy eigenvalues $E_{\nu }=-e^{2}/2a_{0}\nu ^{2}$ and their degeneracies
are also changed accordingly.

Since exact solvability requires polynomial solutions and raising and
lowering operators connecting them, the f\/irst f\/ive types of conf\/inement do
not belong to this category.  Nevertheless, their solutions can be
accurately evaluated for practical applications~\cite{1} and original references
therein.

Concerning the conf\/inement by dihedral angles, the solutions in the other
degrees of freedom maintain their polynomial character in spherical,
spheroidal and parabolic coordinates. In order to establish whether they
belong or not to the exactly solvable category, we must search for the
respective raising and lowering operators.

From the known connection of the solutions of the free hydrogen atom in
parabolic coordinates with the solutions of the harmonic oscillator in
circular coordinates as implied in equation~\eqref{eq3}, we have decided to search f\/irst
for the corresponding operators in the isotropic harmonic oscillator
conf\/ined by dihedral angles, in circular cylindrical and spherical
coordinates.  Correspondingly, in Section~\ref{section2} we identify the eigenfunctions
and energy eigenvalues in the respective coordinates.  In Section~\ref{section3} the
starting point is the familiar raising and lowering operators for the free
harmonic oscillator in cartesian coordinates, followed by the identif\/ication
of their forms in circular cylindrical and spherical coordinates; the next
step is to incorporate the restrictions associated with the eigenfunctions
in the conf\/ining angle, and f\/inally examine its consequences on the other
degrees of freedom. For the latter we make full use of the results of Section~\ref{section2} establishing by mathematical induction the explicit and general forms of
the raising and lowering operators for the circular, spherical radial and
polar angle coordinate eigenfunctions.  In Section~\ref{section4} the relationships
between the sets of degenerate circular cylindrical and spherical
eigenfunctions are explicitly illustrated.  The corresponding results for
the hydrogen atom conf\/ined by dihedral angles are presented in the
successive Sections \ref{section5}--\ref{section7}. The discussion in Section~\ref{section8} revolves around the
superintegrability, exact solvability and special functions for both
conf\/ined-by-dihedral-angle quantum systems, and others.

\section{Eigenfunctions and eigenenergies of the
isotropic\\ harmonic oscillator conf\/ined by dihedral angles}\label{section2}

The Schr\"{o}dinger equation for the harmonic oscillator is known to be
separable in cartesian, circular cylindrical and spherical coordinates
\cite{4, 3}:
\begin{gather}\label{eq7}
x =\rho \cos \varphi =r\sin \vartheta \cos \varphi ,\qquad
 y =\rho \sin \varphi =r\sin \vartheta \sin \varphi ,\qquad
 z =r\cos \vartheta.
\end{gather}
For the system conf\/ined by a dihedral angle def\/ined by the meridian
half-planes $\varphi =0$ and $\varphi =\varphi _{0}$, the boundary conditions
on the respective eigenfunctions are
\begin{gather}
\psi (\rho ,\varphi =0,z) =0, \qquad
 \psi (\rho ,\varphi =\varphi _{0},z) =0,\label{eq8}
\\
\psi (r,\vartheta ,\varphi =0) =0,\qquad
 \psi (r,\vartheta ,\varphi=\varphi _{0}) =0.\label{eq9}
\end{gather}
The separability of the Schr\"{o}dinger equation leads to the common
eigenvalue equation for the square of the $z$-component of the angular
momentum
\begin{gather}\label{eq10}
\frac{d^{2}}{d\varphi ^{2}}\Phi (\varphi )=-\mu ^{2}\Phi (\varphi ).
\end{gather}
As well as the eigenvalue equations for the circular radial and longitudinal
cartesian Hamiltonians
\begin{gather}
\left[ -\frac{\hslash ^{2}}{2m}\left( \frac{1}{\rho }\frac{d}{d\rho }\rho
\frac{d}{d\rho }-\frac{\mu ^{2}}{\rho ^{2}}\right) +\frac{1}{2}m\omega
^{2}\rho ^{2}\right] R(\rho ) =E_{n_{\rho }\mu }R(\rho ),\label{eq11}
\\
\left[ -\frac{\hslash ^{2}}{2m}\frac{d^{2}}{dz^{2}}+\frac{1}{2}m\omega
^{2}z^{2}\right] Z(z) =E_{n_{z}}Z(z),\nonumber%\label{eq12}
\end{gather}
 and the counterparts for the square of the angular momentum and
spherical radial Hamiltonian, respectively,
\begin{gather*}
\left( -\frac{1}{\sin \theta }\frac{d}{d\theta }\sin \theta \frac{d}{d\theta
}+\frac{\mu ^{2}}{\sin ^{2}\theta }\right) \Theta  ( \theta  )
 =\lambda  ( \lambda +1 ) \Theta  ( \theta  ),%\label{eq13}
\\
\left\{ -\frac{\hslash ^{2}}{2m}\left[ \frac{1}{r^{2}}\frac{d}{dr}r^{2}\frac{%
d}{dr}-\frac{\lambda  ( \lambda +1 ) }{r^{2}}\right] +\frac{1}{2}%
m\omega ^{2}r^{2}\right\} R ( \eta  )  =E_{n_{\rho}\lambda}R ( r ).%\label{eq14}
\end{gather*}
The boundary conditions of equations~\eqref{eq8},~\eqref{eq9} lead to the eigenfunctions and
eigenvalues:
\begin{gather}\label{eq15}
\Phi (\varphi )=\sqrt{\frac{2}{\varphi _{0}}}\sin \mu \varphi ,\qquad
\mu =\frac{n_{\varphi }\pi }{\varphi _{0}},\qquad n_{\varphi }=1,2,3,\dots.
\end{gather}
We write f\/irst its circular radial and axial coordinate companion factors in
the cylindrical case, with length units $\sqrt{\hslash /m\omega }$,
\begin{gather}
R_{n_{\rho ,\mu }}(\rho ) =N_{n_{\rho ,\mu }}\rho ^{\mu }e^{-\rho ^{2}/2}
M\left(-n_{\rho },\mu +1,\rho ^{2}\right),\label{eq16}
\\
Z_{n_{z}}(z) =N_{n_{z}}e^{-z^{2}/2} H_{n_{z}}(z) \label{eq17}
\end{gather}
 in terms of singularity removing factors for $\rho \rightarrow 0,$
$\rho \rightarrow \infty $ and $z\rightarrow \infty$,  and the conf\/luent
hypergeometric polynomial of degree $2n_{\rho }$ and the Hermite
polynomials, respectively. The latter correspond to the free harmonic
oscillator in the cartesian axial coordinate, with parity $(-)^{n_{z}}.$ The
radial eigenfunction has the same form as in the free harmonic oscillator
with the integer magnetic quantum number $m$ replaced by the~$\mu $
eigenvalues of equation~\eqref{eq15}. The respective energy eigenvalues in units $\hslash
\omega $ become
\begin{gather}\label{eq18}
E_{n_{\rho }n_{z}\mu }=2n_{\rho }+\mu +n_{z}+\frac{3}{2}.
\end{gather}
Next, we go on to the spherical radial and polar angle coordinate
eigenfunctions:
\begin{gather}
R_{n_{r}n_{_{\theta }}\mu }(r) =N_{n_{r}n_{\theta }\mu }r^{n_{\theta }+\mu
}e^{-r^{2}/2}M\left(-n_{r},n_{\theta }+\mu +\frac{3}{2},r^{2}\right),\label{eq19}
\\
\Theta _{n_{\theta }+\mu }^{\mu }(\cos \theta )  =N_{n_{\theta }+\mu }\frac{\sin ^{\mu}\theta }{2} \left[ _{2}F_{1}\left( -n_{\theta },n_{\theta
}+2\mu +1;\mu +1;\frac{1-\cos \theta }{2}\right) \right.  \notag \\
 \left. \hphantom{\Theta _{n_{\theta }+\mu }^{\mu }(\cos \theta )=}{}
  +(-)^{n_{\theta }}\,{}_{2}F_{1}\left( -n_{\theta },n_{\theta
}+2\mu +1;\mu +1;\frac{1+\cos \theta }{2}\right) \right],\label{eq20}
\end{gather}
where the identif\/ication of the angular momentum label $\lambda
=n_{\theta }+\mu $ is in order. The remarks about the radial eigenfunction
are the same as for its cylindrical counterpart of equation~\eqref{eq16}. Concerning the
polar angle eigenfunction, several remarks are necessary: $i)$~each of the
terms inside the brackets corresponds to a hypergeometric function
representation of the associated Legendre polynomials, $ii)$~for integer
values of $\mu \rightarrow m$ they are equal to each other for $n_{\theta }$
even or odd, containing only even or odd powers of $\cos \theta$, $iii)$~for
the values of $\mu $ in equation~\eqref{eq15}, both contain even and odd powers of $\cos
\theta $ and do not have a well def\/ined parity, $iv)$~their linear combination
in equation~\eqref{eq20} ensures that its parity $(-)^{n_{\theta }}$ is well-def\/ined.
While, the conf\/ining in the dihedral angle breaks the rotational symmetry
around the $z$-axis, the $\mu $ values in each of the terms in equation~\eqref{eq20} also
break the $z\rightarrow -z$ parity; restoration of such a parity requires
the chosen superposition. The corresponding energy eigenvalues are%
\begin{gather}\label{eq21}
E_{n_{r}n_{\theta }\mu }=2n_{r}+n_{\theta }+\mu +\frac{3}{2}.
\end{gather}

For a chosen value of $\mu ,$ the states with $2n_{\rho
}+n_{z}=2n_{r}+n_{\theta }$ share the same energy.  Notice that the
Gaussian exponential factors in equation~\eqref{eq16} and~\eqref{eq17} reproduce the
corresponding factor in equation~\eqref{eq19}; while the $\mu $ powers of $r$ and $\sin
\theta $ in equations~\eqref{eq19} and~\eqref{eq20} reproduce the $\mu $ power of $\rho $ in equation~\eqref{eq16}. Consequently, the eigenfunctions with $n_{\rho }=n_{z}=0$ and $n_{r}=n_{\theta }=0$ are the same in both coordinate systems for all their
common values of $\mu$, or $n_{\varphi }$ for each conf\/ining angle $\varphi_{0}$, equation~\eqref{eq15}.

Tables~\ref{table1}--\ref{table4} illustrate the polynomial factors in the lower excited radial and
axial, and radial and polar angle polynomials equations~\eqref{eq16},~\eqref{eq17} and~\eqref{eq19},~\eqref{eq20} in
the eigenfunctions for the respective coordinate systems. Each one of them
serves as a guide in the identif\/ication of their respective raising and
lowering operators; and the comparison of their products leads to the
relationships between the complete sets of degenerate eigenfunctions in the
alternative coordinates.

\begin{table}[t]\centering
\caption{Polynomial conf\/luent hypergeometric functions.}\label{table1}\vspace{1mm}
\begin{tabular}{|l|c|}
\hline\hline
$n_{\rho}$ & $M\big({-}n_{\rho},\mu+1,\rho^{2}\big)$\bsep{2pt}\tsep{2pt}\\ \hline
0 & 1 \\
1 & $1-\frac{\rho^{2}}{\mu+1}$ \\
2 & $1-\frac{2\rho^{2}}{\mu+1}+\frac{\rho^{4}}{(\mu+1)(\mu+2)}$ \\
3 & $1-\frac{3\rho^{2}}{\mu+1}+\frac{3\rho^{4}}{(\mu+1)(\mu+2)}-\frac{%
\rho^{6}}{(\mu+1)(\mu+2)(\mu+3)}$\bsep{3pt} \\ \hline\hline
\end{tabular}%
\end{table}

\begin{table}[t]\centering
\caption{Hermite polynomials.}\label{table2}\vspace{1mm}
\begin{tabular}{|l|c|}
\hline\hline
$n_z$ & ${\rm Hermite}(n_z,z)$ \\ \hline
0 & 1 \\
1 & $2z$ \\
2 & $-2+4z^{2}$ \\
3 & $-12z+8z^{3}$ \\
4 & $12-48z^{2}+16z^{4}$ \\
5 & $120z-160z^{3}+32z^{5}$ \\
6 & $-120+720z^{2}-480z^{4}+64z^{6}$ \\ \hline\hline
\end{tabular}%
\end{table}

\begin{table}[t]\centering
\caption{Polynomial conf\/luent hypergeometric functions.}\label{table3}\vspace{1mm}
\begin{tabular}{|l|c|}
\hline\hline
$n_r$ & $M\big({-}n_r,\lambda+\frac{3}{2},r^{2}\big)$\tsep{2pt}\bsep{2pt} \\ \hline
0 & 1 \\
1 & $1-\frac{r^{2}}{\lambda+\frac{3}{2}}$ \\
2 & $1-\frac{2r^{2}}{\lambda+\frac{3}{2}}+\frac{r^{4}}{(\lambda+\frac{3}{2}%
)(\lambda+\frac{5}{2})}$ \\
3 & $1-\frac{3r^{2}}{\lambda+\frac{3}{2}}+\frac{3r^{4}}{(\lambda+\frac{3}{2}%
)(\lambda+\frac{5}{2})}-\frac{r^{6}}{(\lambda+\frac{3}{2})(\lambda+\frac{5}{2%
})(\lambda+\frac{7}{2})}$\bsep{6pt} \\ \hline\hline
\end{tabular}%
\end{table}
\begin{table}[t]\centering
\caption{Associated Legendre polynomials with def\/inite parity $(-)^{n_{\theta}}$.}\label{table4}\vspace{1mm}
\begin{tabular}{|l|c|}
\hline\hline
$n_{\theta }$ & Even or odd powers of$_{\text{ }2}F_{1}\big({-}n_{\theta
},n_{\theta }+2\mu +1;\mu +1;\frac{1-\cos \theta }{2}\big)$ \tsep{2pt}\bsep{2pt} \\ \hline
0 & 1 \\
1 & $\cos \theta $ \\
2 & $\frac{-1+(2\mu +3){\cos ^{2}\theta }}{2(\mu +1)}$ \\
3 & $\frac{-3\cos \theta +(2\mu +5){\cos ^{3}\theta }}{2(\mu +1)}$ \\
4 & $\frac{3-6(2\mu +5){\cos ^{2}}\theta +(2\mu +5)(2\mu +7){\cos ^{4}\theta
}}{4(\mu +1)(\mu +2)}$ \\
5 & $\frac{15\cos \theta -10(2\mu +7){\cos ^{3}}\theta +(2\mu +7)(2\mu +9){%
\cos ^{5}\theta }}{4(\mu +1)(\mu +2)}$ \\
6 & $\frac{-15+45((2\mu +7)\cos ^{2}\theta -15(2\mu +7)(2\mu +9){\cos ^{4}}%
\theta +(2\mu +7)(2\mu +9)(2\mu +11){\cos ^{6}\theta }}{8(\mu +1)(\mu
+2)(\mu +3)}$\bsep{4pt} \\ \hline\hline
\end{tabular}
\end{table}

\section{Raising and lowering operators\\ for the successive degrees of freedom}\label{section3}

The creation and annihilation operators for the free harmonic oscillator in
each cartesian coordinate constitute our starting point and immediate bridge
to their circular cylindrical and spherical counterparts:
\begin{gather}
 \widehat{i}\left(x\pm \frac{\partial }{\partial x}\right)+\widehat{j}\left(y\pm \frac{\partial }{\partial y}\right)+\widehat{k}\left(z\pm \frac{\partial }{\partial z}\right)
=\big(\vec{\rho }\pm \nabla _{\vec{\rho }}\big)+\widehat{k}
\left(z\pm \frac{\partial }{\partial z}\right)=\big(\vec{r}\pm \nabla _{\vec{r}}\big).\label{eq22}
\end{gather}

Let us consider the circular coordinate components rewritten in terms of the
cartesian unit vectors $\widehat{i}$ and $\widehat{j}$:
\begin{gather}
\big(\vec{\rho }\pm \nabla _{\vec{\rho }}\big)  = \widehat{\rho }\left(\rho \pm \frac{\partial }{\partial \rho }\right)\pm \frac{\widehat{\varphi }
}{\rho }\frac{\partial }{\partial \varphi }\nonumber\\
\hphantom{\big(\vec{\rho }\pm \nabla _{\vec{\rho }}\big)}{}
 = \big(\,\widehat{i}\cos \varphi +\widehat{j}\sin \varphi \big)
 \left(\rho \pm \frac{\partial }{\partial \rho }\right)\pm \big({-}\widehat{i}\sin \varphi +\widehat{j}\cos
\varphi \big)\frac{1}{\rho }\frac{\partial }{\partial \varphi }.\label{eq23}
\end{gather}

The application of the operators in the $\widehat{i}$ direction on the
eigenfunctions of equation~\eqref{eq15}, with the idea of raising and lowering the
quantum number $n_{\varphi }$, suggests the change of~$\varphi $ by~$\pi
\varphi /\varphi _{0}$ in the respective trigonometric functions. In
fact, for the angular factor of the radial operator term the result is
\begin{gather}\label{eq24}
\cos \frac{\pi \varphi }{\varphi _{0}}\sin \frac{n_{\varphi }\pi \varphi }{\varphi _{0}}
=\frac{1}{2}\left[ \sin \frac{(n_{\varphi }+1)\pi \varphi }{\varphi _{0}}+\sin \frac{(n_{\varphi }-1)\pi \varphi }{\varphi _{0}}\right]
\end{gather}
achieving the desired raising and lowering actions in the
eigenfunctions. For the angular derivative term, the corresponding result
is similar,
\begin{gather}
-\sin \frac{\pi \varphi }{\varphi _{0}}\frac{d}{d\varphi }\sin \frac{n_{\varphi }\pi \varphi }{\varphi _{0}}  = -\frac{n_{\varphi }\pi }{\varphi
_{0}}\sin \frac{\pi \varphi }{\varphi _{0}}\cos \frac{n_{\varphi }\pi
\varphi }{\varphi _{0}}  \notag \\
\hphantom{-\sin \frac{\pi \varphi }{\varphi _{0}}\frac{d}{d\varphi }\sin \frac{n_{\varphi }\pi \varphi }{\varphi _{0}}}{}
= -\frac{n_{\varphi }\pi }{\varphi _{0}}\frac{1}{2}\left[ \sin \frac{(n_{\varphi }+1)\pi \varphi }{\varphi _{0}}-\sin \frac{(n_{\varphi }-1)\pi
\varphi }{\varphi _{0}}\right] .\label{eq25}
\end{gather}

The appropriate linear combinations of equations \eqref{eq24} and \eqref{eq25} allow the
identif\/ication of the respective raising and lowering operators:
\begin{gather}
\left[ \cos \frac{\pi \varphi }{\varphi }+\frac{\varphi _{0}}{n_{\varphi
}\pi }\sin \frac{\pi \varphi }{\varphi _{0}}\frac{d}{d\varphi }\right] \sin
\frac{n_{\varphi }\pi \varphi }{\varphi _{0}} =\sin \frac{(n_{\varphi }+1)\pi
\varphi }{\varphi _{0}},\label{eq26}
\\
\left[ \cos \frac{\pi \varphi }{\varphi _{0}}-\frac{\varphi _{0}}{n_{\varphi
}\pi }\sin \frac{\pi \varphi }{\varphi _{0}}\frac{d}{d\varphi }\right] \sin
\frac{n_{\varphi }\pi \varphi }{\varphi _{0}} =\sin \frac{(n_{\varphi }-1)\pi
\varphi }{\varphi _{0}}.\label{eq27}
\end{gather}

They have the same structure as the corresponding operators for the particle
conf\/ined in the one-dimensional inf\/inite square well~\cite{5}, and are also
connected with the two-dimensional harmonic oscillator in a sector
operators~\cite{6}.

The reader may also examine the actions of the corresponding operators in
the~$\widehat{j}$ direction recognizing that they lead to the companion
cosine functions satisfying the Neumann boundary condition instead of the
Dirichlet one of equations~\eqref{eq8},~\eqref{eq9}.

For a f\/ixed value of $\mu$, the radial dependent operators in the
combinations
\begin{gather*}%\label{eq28}
\rho \pm \frac{d}{d\rho }\mp \frac{\mu }{\rho }
\end{gather*}
guarantee the removal of the singularities at $\rho \rightarrow 0$
and $\rho \rightarrow \infty ,$ including the annihilation of the
eigenstates $\rho ^{\mu }e^{-\rho ^{2}/2}$ when the upper signs are chosen.
 For the states with radial excitations, we adapt the recurrence relations
for the conf\/luent hypergeometric functions (13.4.10) and (13.4.11) in~\cite{7}:
\begin{gather}
\left(a+z\frac{d}{dz}\right)M(a,b,z) =aM(a+1,b,z),\label{eq29}
\\
\left(b-a-z+z\frac{d}{dz}\right) M(a,b,z) =(b-a)M(a-1,b,z)\label{eq30}
\end{gather}
with the correspondences $a=-n_{\rho }$, $b=\mu +1$ and $z=\rho
^{2}$ in equation~\eqref{eq16}. The f\/irst one with $a+1=-(n_{\rho }-1)$is the lowering
operator, while the second one with $a-1=-(n_{\rho }+1)$ is the raising
operator. The interested readers can check these actions between the
consecutive entries in Table~\ref{table1}.

Since the harmonic oscillator in the $z$-direction is free, the actions of the
corresponding operators in equation~\eqref{eq22} on the Hermite polynomials of Table~\ref{table2}
are the familiar ones.

In the spherical coordinates, the raising and lowering operators in the $\varphi $ angle are those in equations~\eqref{eq26},~\eqref{eq27}, while those in the radial
coordinate are also given by equations~\eqref{eq29},~\eqref{eq30} with the correspondences $a=-n_{r}$, $b=\lambda +\frac{3}{2},$ and $z=r^{2}$ in equation~\eqref{eq19}. Their
actions can be verif\/ied by applying them in the entries of Table~\ref{table3}. For
the polar angle the recurrence relations for the associated Legendre
functions (8.5.3), (8.5.4) in~\cite{7}
\begin{gather}
\left[ \lambda z-\big(z^{2}-1\big)\frac{d}{dz}\right] P_{\lambda }^{\mu
}(z) =(\lambda +\mu )P_{\lambda -1}^{\mu }(z),\label{eq31}
\\
\left[ (\lambda +1)z+\big(\lambda ^{2}-1\big)\frac{d}{dz}\right] P_{\lambda }^{\mu
}(z) =(\lambda -\mu +1)P_{\lambda +1}^{\mu }(z)\label{eq32}
\end{gather}
 can not be applied directly to the eigenfunctions in equation~\eqref{eq20} due
to some of the remarks in that paragraph.

Nevertheless, equations~\eqref{eq31},~\eqref{eq32} suggest the structure of the correct lowering
and raising operators for the polar angle eigenfunctions with parity $\pi
=(-)^{n_{\theta }}$:
\begin{gather}
 \left[ A_{-}^{(n_{\theta })}z+\big(B_{-}^{(n_{\theta
})}z^{2}+C_{-}^{(n_{\theta })}\big)\frac{d}{dz}\right] \left. _{\text{\ }%
2}F_{1}\left( -n_{\theta },2\mu +n_{\theta }+1;\mu +1;\frac{1-z}{2}\right)
\right\vert _{\pi =(-1)^{n_{\theta }}}  \notag \\
\left. \qquad= {}_{2}F_{1}\left( -(n_{\theta }-1),2\mu +n_{\theta };\mu +1;\frac{1-z%
}{2}\right) \right\vert _{\pi =(-1)^{n_{\theta -1}}},\label{eq33}
\\
 \left[ A_{+}^{(n_{\theta })}z+\big(B_{+}^{(n_{\theta
})}z^{2}+C_{+}^{(n_{\theta })}\big)\frac{d}{dz}\right] \left. _{2}F_{1}\left(
-n_{\theta },2\mu +n_{\theta }+1;\mu +1;\frac{1-z}{2}\right) \right\vert
_{\pi =(-)^{n_{\theta }}}  \notag \\
\left. \qquad={} _{2}F_{1}\left( -(n_{\theta }+1),2\mu +n_{\theta }+2;\mu +1;\frac{%
1-z}{2}\right) \right\vert _{\pi =(-)^{n_{\theta +1}}}.\label{eq34}
\end{gather}

Their applications to the entries in Table~\ref{table4} lead by mathematical induction
to the general results:
\begin{alignat}{4}
& A_{-}^{(n_{\theta })} =1,\qquad &&
 B_{-}^{(n_{\theta })} =-\frac{1}{n_{\theta }},\qquad &&
 C_{-}^{(n_{\theta })} =\frac{1}{n_{\theta }}, & \label{eq35}
\\
& A_{+}^{(n_{\theta })} =1,\qquad &&
 B_{+}^{(n_{\theta })} =\frac{1}{2\mu+n_{\theta }+1},\qquad &&
 C_{+}^{(n_{\theta })} =-\frac{1}{2\mu +n_{\theta }+1}, &\label{eq36}
\end{alignat}
and in particular, $A_{-}^{(0)}=0$ for annihilation of the eigenfunction
with $n_{\theta }=0$, which is~$1$ and also has a vanishing derivative,
making the values of $B_{-}^{(0)}$ and $C_{-}^{(0)}$ irrelevant; also \mbox{$A_{+}^{(0)}=1$}, and~$B_{+}^{(0)}$ and~$C_{+}^{(0)}$ irrelevant. Notice also
the equal magnitudes and opposite signs of the respective~$B$ and~$C$
coef\/f\/icients leading to the common factors $\mp (z^{2}-1)$ of the derivative
terms in equations~\eqref{eq33} and~\eqref{eq34}, respectively. The values of $A_{-}^{(n_{\theta
})}$ and $B_{-}^{(n_{\theta })}$ guarantee the vanishing of the term with
the $n_{\theta }$ power of~$z$, while the values of $A_{+}^{(n_{\theta })}$
and $B_{+}^{(n_{\theta })}$ guarantee the highest power $(n_{\theta }+1)$ of~$z$ and its correct coef\/f\/icient in the r.h.s.\ eigenfunction in equation~\eqref{eq34}.

The energy eigenvalues in equations~\eqref{eq18} and~\eqref{eq21} contain the same zero-point,
radial and axial or polar angle contributions as in the free harmonic
oscillator. The dif\/ference for the oscillator conf\/ined by dihedral angles
corresponds to the~$\mu $ contribution which according to equation~\eqref{eq15} is
quantized with integer values of~$n_{\varphi }$ and spacings inversely
proportional to the conf\/ining angle~$\varphi _{0}$.

Let us start with the case of $\varphi _{0}=\pi$, for which the dihedral
angle becomes a plane and~$\mu $ becomes $n_{\varphi }$; consequently, its
contributions to the energy are on the same footing as those of the other
degrees of freedom. Next, let us consider the conf\/ining angles in the
interval $\pi \leqslant \varphi _{0}\leqslant 2\pi $ for which the~$\mu $
energy contribution is less than those for the other degrees of freedom,
taking its lowest value of one half for $\varphi _{0}=2\pi$; this is the
quasi-free limit for the harmonic oscillator, being excluded from the $xz$
half plane $(0<x<\infty ,y=0,-\infty <z<\infty )$, and def\/initely dif\/ferent
from the free harmonic oscillator. On the other hand, for the interval $\pi
\geqslant \varphi _{0}>0,$ as the dihedral angles is closing starting from
the plane, the~$\mu $ contribution to the energy becomes larger than the
contributions from the other degrees of freedom, tending to inf\/inity as $\varphi _{0}\rightarrow 0$ when the moving meridian half-plane of the
conf\/ining angle approaches the position of the~$xz$ half plane.

\section{Interbasis expansions of degenerate eigenfunctions\\ in circular
cylindrical and spherical coordinates}\label{section4}

In this section the complete eigenfunctions of the harmonic oscillator
conf\/ined by dihedral angles in circular cylindrical and spherical
coordinates are compared, illustrating explicitly the relationships of their
sets with the same energy, as counterparts of \cite{10,9,8} for the free
oscillator.

The complete eigenfunctions are the products of the common factor of equation~\eqref{eq15}, and the respective companions of equations~\eqref{eq16},~\eqref{eq17} and equations~\eqref{eq19},~\eqref{eq20}:
\begin{gather*}
\psi _{n_{\rho }\mu n_{z}}(\rho ,\varphi ,z) =R_{n_{\rho }\mu }(\rho )\Phi
_{\mu }(\varphi )Z_{n_{z}}(z),%\label{eq37}
\qquad
\psi _{n_{r}n_{\theta }\mu }(r,\theta ,\varphi ) =R_{n_{r}n_{\theta }\mu }(r)
 \Theta _{n_{\theta }+\mu }^{\mu }(\theta )\Phi _{\mu }(\varphi ).%\label{eq38}
\end{gather*}

For a chosen value of $\mu$, the eigenstates without excitations in the
other degrees of freedom are the same in both coordinate systems, which is
expressed as
\begin{gather*}%\label{eq39}
\left\vert n_{\rho }=0,\mu ,n_{z}=0 \right\rangle
=
\left\vert n_{r}=0, n_{\theta }=0,\mu \right\rangle
\end{gather*}
using the ket notation. The eigenstates with excitations in the
other degrees of freedom share the common factors in equations~\eqref{eq16},~\eqref{eq17} and
\eqref{eq19},~\eqref{eq20} as already discussed in the paragraph after equation~\eqref{eq21}.
Correspondingly, their comparison can be restricted to that of the products
of the polynomials in Tables~\ref{table1} and~\ref{table2} and in Tables~\ref{table3} and~\ref{table4}, respectively.
Such a comparison will be implemented for the eigenstates involving the
successive rows in those tables, and making use of the connections between
the respective coordinates, equation~\eqref{eq7}.

The lowest excitation eigenstates, with negative $z$-parity and energy $(\mu
+5/2)$ are also the same:
\begin{gather*}%\label{eq40}
\left\vert n_{\rho }=0, \mu ,n_{z}=1 \right\rangle=\left\vert n_{r}=0,
n_{\theta }=1,\mu \right\rangle
\end{gather*}
since $z=r\cos \theta$.

The next sets of excited states with positive $z$-parity and energy $(\mu +
\frac{7}{2})$, $\left\vert n_{\rho }=1, \mu ,n_{z}=0\right\rangle$, $\left\vert n_{\rho }= 0,\mu ,n_{z}=2 \right\rangle$ and $\left\vert n_{r}
=1,n_{\theta }=0,\mu \right\rangle$, $ \left\vert n_{r} =0,n_{\theta }=2,\mu \right\rangle$
 involve the products of the respective polynomials:
\begin{alignat*}{3}
& \left( 1-\frac{1}{\mu +1}\rho ^{2}\right) \cdot 1,\qquad && 1\cdot \left(
2z^{2}-1\right), & \\ %\label{eq41}\\
& \left( 1-\frac{1}{\mu +\frac{3}{2}}r^{2}\right) \cdot 1,\qquad && r^{2}\cdot
\frac{(2\mu +3)\cos ^{2}\theta -1}{2(\mu +1)}. & %\label{eq42}
\end{alignat*}

By using $r^{2}\cos ^{2}\theta =z^{2}$ and $r^{2}=\rho ^{2}+z^{2}$, the last
pair can be written as the linear combinations of the f\/irst pair:
\begin{gather}
\left( 1-\frac{2(\rho ^{2}+z^{2})}{2\mu +3}\right)  =\frac{2(\mu +1)}{2\mu +3}
\left( 1-\frac{\rho ^{2}}{\mu +1}\right) -\frac{1}{2\mu +3}\big(2z^{2}-1\big),\label{eq43}
\\
\frac{(2\mu +3)z^{2}-\rho ^{2}-z^{2})}{2(\mu +1)}  =\frac{1}{2\mu +3}\left( 1-
\frac{\rho ^{2}}{\mu +1}\right) +\frac{2(\mu +1)}{2\mu +3}\big(2z^{2}-1\big).\label{eq44}
\end{gather}

The relationships between the normalized eigenfunctions follows from these
relationships between the polynomials
\begin{gather*}%\label{eq45}
\begin{pmatrix}
 \vert n_{r}=1,  n_{\theta }=0,\mu \rangle \\
\\
 \vert n_{r}=0,  n_{\theta }=2,\mu \rangle%
\end{pmatrix}
=
\begin{pmatrix}
\sqrt{\frac{2(\mu +1)}{2\mu +3}}\ -\sqrt{\frac{1}{2\mu +3}} \\
\sqrt{\frac{1}{2\mu +3}}\ +\sqrt{\frac{2(\mu +1)}{2\mu +3}}%
\end{pmatrix}
\begin{pmatrix}
 \vert n_{\rho }=1,  \mu ,n_{z}=0\rangle \\
\\
 \vert n_{\rho }=0,  \mu ,n_{z}=2\rangle
\end{pmatrix}.
\end{gather*}

In the same way we have also obtained the following results for the
successive eigenstates with higher excitations and energies  $E=\mu +\frac{9}{2},\mu +\frac{11}{2},\mu +\frac{13}{2}$, respectively,
\begin{gather*}
\begin{pmatrix}
|n_{r}=1,n_{\theta }=1,\mu \rangle \\
\\
|n_{r}=0,n_{\theta }=3,\mu \rangle%
\end{pmatrix}
\begin{pmatrix}
\sqrt{\frac{2(\mu +1)}{2\mu +5}} & -\sqrt{\frac{3}{2\mu +5}} \\
\sqrt{\frac{3}{2\mu +5}} & \sqrt{\frac{2(\mu +1)}{2\mu +5}}%
\end{pmatrix}
\begin{pmatrix}
|n_{\rho }=1,\mu ,n_{z}=1\rangle \\
\\
|n_{\rho }=0,\mu ,n_{z}=3\rangle%
\end{pmatrix},%\label{eq46}
\\
\begin{pmatrix}
|20\mu \rangle \\
\\
|12\mu \rangle \\
\\
|04\mu \rangle
\end{pmatrix}
=
\begin{pmatrix}
\sqrt{\frac{4(\mu +1)(\mu +2)}{(2\mu +3)(2\mu +5)}} & -\sqrt{\frac{4(\mu +1)%
}{(2\mu +3)(2\mu +5)}} & -\sqrt{\frac{3}{(2\mu +3)(2\mu +5)}} \\
\sqrt{\frac{4(\mu +2)}{(2\mu +3)(2\mu +7)}} & \frac{2\mu +1}{\sqrt{(2\mu
+3)(2\mu +7)}} & \sqrt{\frac{12(\mu +1)}{(2\mu +3)(2\mu +7)}} \\
\sqrt{\frac{3}{(2\mu +5)(2\mu +7)}} & \sqrt{\frac{12(\mu +2)}{(2\mu +5)(2\mu
+7)}} & -\sqrt{\frac{4(\mu +1)(\mu +2)}{(2\mu +5)(2\mu +7)}}%
\end{pmatrix}
\begin{pmatrix}
|2\mu 0\rangle \\
\\
|1\mu 2\rangle \\
\\
|0\mu 4\rangle%
\end{pmatrix},%\label{eq47}
\\
\begin{pmatrix}
|21\mu \rangle \\
\\
|13\mu \rangle \\
\\
|05\mu \rangle%
\end{pmatrix}%
=
\begin{pmatrix}
\sqrt{\frac{4(\mu +1)(\mu +2)}{(2\mu +5)(2\mu +7)}} & -\sqrt{\frac{12(\mu +1)%
}{(2\mu +5)(2\mu +7)}} & \sqrt{\frac{15}{(2\mu +5)(2\mu +7)}} \\
\sqrt{\frac{12(\mu +2)}{(2\mu +5)(2\mu +9)}} & \frac{2\mu -1}{\sqrt{(2\mu
+5)(2\mu +9)}} & -\sqrt{\frac{20(\mu +1)}{(2\mu +5)(2\mu +9)}} \\
\sqrt{\frac{15}{(2\mu +7)(2\mu +9)}} & \sqrt{\frac{20(\mu +2)}{(2\mu
+7)(2\mu +9)}} & \sqrt{\frac{4(\mu +1)(\mu +2)}{(2\mu +7)(2\mu +9)}}%
\end{pmatrix}
\begin{pmatrix}
|2\mu 1\rangle \\
\\
|1\mu 3\rangle \\
\\
|0\mu 5\rangle
\end{pmatrix}.%\label{eq48}
\end{gather*}

Notice the orthonormality of the transformation matrices, implying the
Solomonic distribution of the coef\/f\/icients in equations~\eqref{eq43},~\eqref{eq44} and their
counterparts between the entries in the matrices and the normalization
constants of the respective eigenfunctions. The sets of eigenstates come
by pairs of singlets, doublets, triplets, and in general multiplets of order
$(2n_{\rho}+n_{z}^{e}+2)/2=(2n_{\rho }+n_{z}^{0}+1)/2$  with $n_{z}^{0}=n_{z}^{e}+1$, etc., of positive and negative $z$-parity, and with
equal successive energy spacings of $\hslash \omega $, for f\/ixed values of~$\mu$. The relationships in this section are valid for any chosen values of $\mu $ from equation~\eqref{eq15}, for any~$n_{\varphi }$ excitation and for each angle of
conf\/inement~$\varphi _{0}$.

\section{Eigenstates and eigenenergies of the hydrogen atom\\ conf\/ined by
dihedral angles}\label{section5}

The Schr\"{o}dinger equation for the hydrogen atom is separable in spherical
coordinates, equations~\eqref{eq7}, as well as in parabolic and prolate spheroidal
coordinates \cite{15-1,15-2,13,16,12,18, 11,14,17}:
\begin{gather}
x  = \xi \eta \cos \varphi =f\sqrt{(u^{2}-1)(1-v^{2})}\cos \varphi,  \qquad
y  = \xi \eta \sin \varphi =f\sqrt{(u^{2}-1)(1-v^{2})}\sin \varphi , \notag \\
z  = \frac{\xi ^{2}-\eta ^{2}}{2}=fuv.\label{eq49}
\end{gather}

The eigenfunctions of the hydrogen atom conf\/ined by dihedral angles in the
respective coordinates must satisfy the equations~\eqref{eq9} and their counterparts
\begin{alignat}{3}
& \psi ^{P}(\xi ,\eta ,\varphi =0)=0,\qquad && \psi ^{P}(\xi ,\eta ,\varphi =\varphi
_{0}) =0,& \label{eq50}
\\
& \psi ^{PS}(u,v,\varphi =0)=0, \qquad && \psi ^{PS}(u,v,\varphi =\varphi _{0}) =0.& \label{eq51}
\end{alignat}

They also share the eigenvalue equation and its solutions in equations~\eqref{eq10},~\eqref{eq15}.
In spherical coordinates, the eigenvalue equation for the square of the
angular momentum is also the same as in Section~\ref{section2}, and for the spherical radial Hamiltonian:
\begin{gather*}
\left\{ -\frac{\hslash ^{2}}{2m_{e}}\left[ \frac{1}{r^{2}}\frac{d}{dr}r^{2}
\frac{d}{dr}-\frac{\lambda \left( \lambda +1\right) }{r^{2}}\right] -\frac{
e^{2}}{r}\right\} R(r)=ER\left( r\right) .%\label{eq52}
\end{gather*}

In parabolic coordinates, the $z$-component of the Runge--Lenz vector is the
constant of motion leading to the separated ordinary dif\/ferential equations:%
\begin{gather}
\left[ -\frac{\hslash ^{2}}{2m_{e}}\left( \frac{1}{\xi }\frac{d}{d\xi }\xi
\frac{d}{d\xi }-\frac{\mu ^{2}}{\xi ^{2}}\right) -E\xi ^{2}\right] \Xi
\left( \xi \right)  =A_{\xi }\Xi \left( \xi \right) ,\label{eq53}
\\
\left[ -\frac{\hslash ^{2}}{2m_{e}}\left( \frac{1}{\eta }\frac{d}{d\eta }
\eta \frac{d}{d\eta }-\frac{\mu ^{2}}{\eta ^{2}}\right) -E\eta ^{2}\right]
H\left( \eta \right)  =A_{\eta }H\left( \eta \right) ,\label{eq54}
\end{gather}
where the separation constants must satisfy the condition%
\begin{gather}\label{eq55}
A_{\xi }+A_{\eta }=2e^{2}.
\end{gather}
For the bound states with negative energy of the hydrogen atom, these
equations are equivalent to that of the two-dimensional harmonic oscillator
in two dimensions in the circular radial coordinate, equation~\eqref{eq11}.

The constant of motion in prolate spheroidal coordinates is a linear
combination of the square of the angular momentum, the $z$-component of the
Runge--Lenz vector and the Hamiltonian~\cite{1}, leading to the separated ordinary
dif\/ferential equations%
\begin{gather*}
\left\{ -\left[ \frac{d}{du}\left( u^{2}-1\right) \frac{d}{du}-\frac{\mu ^{2}
}{u^{2}-1}\right] -\frac{2me^{2}fu}{\hslash ^{2}}-\frac{2me^{2}Ef^{2}u^{2}}{\hslash ^{2}}\right\} U\left( u\right)  =-AU\left( u\right) ,%\label{eq56}
\\
\left\{ -\left[ \frac{d}{dv}\left( 1-v^{2}\right) \frac{d}{dv}-\frac{\mu ^{2}%
}{1-v^{2}}\right] +\frac{2me^{2}fv}{\hslash ^{2}}+\frac{2me^{2}Ef^{2}v^{2}}{%
\hslash ^{2}}\right\} V\left( v\right)  =AV\left( v\right) .%\label{eq57}
\end{gather*}
These two equations are the same in the respective domains $1\leqslant
u<\infty \leqslant $ and $-1\leqslant v\leqslant 1$.

Their remaining radial and angular solutions have the same forms
as those of the free hydrogen with the substitution of the integer magnetic
quantum number $m$ with the eigenvalue $\mu $ in equation~\eqref{eq15}.

In spherical coordinates, the polar angle eigenfunction of the square
angular momentum is the one in equation~\eqref{eq20}, and the radial eigenfunction of the
energy becomes
\begin{gather}\label{eq58}
R_{n_{r}\lambda }^{S}(r)=N_{n_{r}\lambda }r^{\lambda }e^{-r/\nu
a_{0}}M(-n_{r},2\lambda +2,2r/\nu a_{0}),
\end{gather}
with $\lambda =n_{\theta }+\mu$. The energy eigenvalue $%
E=-e^{2}/2a_{0}\nu ^{2}$ involves the non-integer principal quantum label%
\begin{gather}\label{eq59}
\nu =n_{r}+n_{\theta }+\mu +1.
\end{gather}

{\sloppy In parabolic coordinates, by adopting the harmonic oscillator
parametrization in equa\-tions~\eqref{eq11} and~\eqref{eq18}
\begin{gather*}
E  =-\frac{1}{2}m\omega ^{2}=-\frac{e^{2}}{2\nu ^{2}a_{0}},%\label{eq60}
\qquad
A_{\xi } =\hslash \omega  ( 2n_{\xi }+\mu +1 ),%\label{eq61}
\qquad
A_{\eta } =\hslash \omega  ( 2n_{\eta }+\mu +1 ) ,%\label{eq62}
\end{gather*}
the identif\/ication of their frequency $\omega =e^{2}/\hslash \nu $, length
unit $\hslash /m\omega =\nu a_{0}$, and the connection between the quantum
labels follow immediately:
\begin{gather}\label{eq63}
\nu =n_{\xi }+n_{\eta }+\mu +1,
\end{gather}
as the counterpart of equation~\eqref{eq59}, and consistent with equation~\eqref{eq55}.

}

The respective eigenfunctions share the same forms
\begin{gather}
\Xi _{n_{_{\xi }}\mu } =N_{n_{\xi }\mu } \xi ^{\mu } e^{-\xi
^{2}/2\nu a_{0}} M\big({-}n_{\xi },\mu +1,\xi ^{2}/\nu a_{0}\big),\label{eq64}
\\
H_{n_{\eta }\mu } =N_{n_{\eta }\mu }\eta ^{\mu } e^{-\eta ^{2}/2\nu
a_{0}} M\big({-}n_{\eta },\mu +1,\eta ^{2}/\nu a_{0}\big),\label{eq65}
\end{gather}
which coincide with that of equation~\eqref{eq16} for the harmonic oscillator in the
circular radial coordinate. Notice also that $\xi \eta $ in equations~\eqref{eq49}
coincides with $\rho =r\sin \theta $ in equations~\eqref{eq7}, and that $(\xi ^{2}+\eta
^{2})/2=r$ using equations~\eqref{eq49} ensuring that the products of the factors in equations~\eqref{eq64},~\eqref{eq65} and equations~\eqref{eq58},~\eqref{eq20} share the same factors removing the singularities
at the $z$-axis and at inf\/inity.

In prolate spheroidal coordinates the spheroidal and hyperboloidal
eigenfunctions also share the same dif\/ferential equation and forms:
\begin{gather}
U_{n_{u}\mu }(u) =N_{n_{u}\mu }\big(u^{2}-1\big)^{\mu /2}e^{-fu/\nu
a_{0}}S_{n_{u}}^{\mu }(u),\label{eq66}
\\
V_{n_{v}\mu }(v) =N_{n_{v}\mu }\big(1-v^{2}\big)^{\mu /2}e^{-fv/\nu
a_{0}}S_{n_{v}}^{\mu }(v),\label{eq67}
\end{gather}
with polynomials
\begin{gather}\label{eq68}
S_{n_{u}}^{\mu }(u)=\sum^{n_{u}+n_v}_{s=0}c_{s}(u-1)^{s}, \qquad
 S_{n_{v}}^{\mu }(v)=\sum^{n_{u}+n_{v}}_{s=0}
c_{s}(v-1)^{s},
\end{gather}
with common coef\/f\/icients satisfying three-term recurrence relations \cite{16, 1}
\begin{gather}
\frac{2f}{a_{0}\nu }(n_{u}+n_{v}+1-s)c_{s-1}+\left[ s\left( s+2\mu +1-\frac{2f}{a_{0}\nu }\right) -A\right] c_{s}\nonumber\\
\qquad{}+2(s+1)(s+\mu +1)c_{s+1}=0,\label{eq69}
\end{gather}%
and in dif\/ferent ranges of their variables $1\leqslant u<\infty$, $-1\leqslant v\leqslant 1$. Now, equations~\eqref{eq49} and~\eqref{eq7} lead to the
identif\/ication of $f\sqrt{(u^{2}-1)(1-v^{2})}=r_{1}\sin \theta _{1}$
and $f(u+v)=r_{1}$, where the subindex~1 indicates that the spherical
coordinates are referred to the position of the nucleus as the origin and
located at the focus ($x=0$, $y=0$, $z=-f$). Correspon\-dingly, the
pro\-ducts of equations~\eqref{eq66},~\eqref{eq67} and equations~\eqref{eq58},~\eqref{eq20} also share the same factors
removing the singularities at the $z$-axis and at inf\/inity.

The principal quantum label becomes
\begin{gather}\label{eq70}
\nu =n_{u}+n_{v}+\mu +1.
\end{gather}

Comparison of equations~\eqref{eq59}, \eqref{eq63} and~\eqref{eq70} with a common value of~$\mu$ lead us
to recognize the common value of their energy for the states with
\begin{gather*}%\label{eq71}
n_{r}+n_{\theta }=n_{\xi }+n_{\eta }=n_{u}+n_{v},
\end{gather*}
and with a degeneracy one unit above such sums.

Table~\ref{table5} illustrates the explicit expressions of the polynomial conf\/luent
hypergeometric functions of equation~\eqref{eq58}. Their polar angle companions are those
in Table~\ref{table4}.

The parabolic coordinate conf\/luent hypergeometric polynomials of equations~\eqref{eq64},~\eqref{eq65} coin\-cide with those in Table~\ref{table1} with the respective replacement of~$\rho $ with~$\xi $ and~$\eta $, as already discussed.

\begin{table}[t]\centering
\caption{Conf\/luent hypergeometric polynomials.}\label{table5}\vspace{1mm}
\begin{tabular}{|l|c|}
\hline\hline
$n_{r}$ & $M(-n_{r},2\lambda +2,2r)$\bsep{1pt}\tsep{1pt} \\ \hline
0 & 1 \\
1 & $1-\frac{2r}{2\lambda +2}$ \\
2 & $1-\frac{2(2r)}{2\lambda +2}+\frac{(2r)^{2}}{(2\lambda +2)(2\lambda +3)}$
\\
3 & $1-\frac{3(2r)}{2\lambda +2}+\frac{3(2r)^{2}}{(2\lambda +2)(2\lambda +3)}%
-\frac{(2r)^{3}}{(2\lambda +2)(2\lambda +3)(2\lambda +4)}$\bsep{3pt} \\ \hline\hline
\end{tabular}
\end{table}

\begin{table}[t]\centering
\caption{The ratios of normalization constants $N_{n_{1}n_{2}\mu}//N_{00\mu}$ and polynomial factors for excited states.}\label{table6}
\vspace{1mm}
\begin{tabular}{|l|c|c|}
\hline\hline
States\bsep{1pt}\tsep{1pt} & Spherical coordinates & Parabolic coordinates \\ \hline
01$\mu $ & $\frac{2}{\sqrt{2+\mu }}\frac{r\cos \theta }{a_{0}\nu }$ & $%
\tfrac{1+\mu }{\sqrt{2+\mu }}\left( 1-\tfrac{\eta ^{2}}{a_{0}\nu \left(
1+\mu \right) }\right) $ \\
10$\mu $ & $\frac{2\left( 1+\mu \right) }{\sqrt{2+\mu }}\left( 1-\frac{r}{%
a_{0}\nu \left( 1+\mu \right) }\right) $ & $\tfrac{1+\mu }{\sqrt{2+\mu }}%
\left( 1-\tfrac{\xi ^{2}}{a_{0}\nu \left( 1+\mu \right) }\right) $ \\
02$\mu $ & $\sqrt{\frac{4\left( 1+\mu \right) }{\left( 2+\mu \right) \left(
3+\mu \right) \left( 3+2\mu \right) }}\tfrac{r^{2}\left[ -1+\left( 3+2\mu
\right) \cos ^{2}\theta \right] }{\left( a_{0}\nu \right) ^{2}\left( 1+\mu
\right) }$ & $\tfrac{\left( 1+\mu \right) \sqrt{2+\mu }}{\sqrt{3+\mu }}%
\left( 1-\tfrac{2\eta ^{2}}{a_{0}\nu \left( 1+\mu \right) }+\tfrac{\eta ^{4}%
}{a_{0}^{2}\nu ^{2}\left( 2+\mu \right) \left( 1+\mu \right) }\right) $ \\
11$\mu $ & $\tfrac{\sqrt{8}r\cos \theta }{a_{0}\nu }\left( 1-\tfrac{r}{%
a_{0}\nu \left( 2+\mu \right) }\right) $ & $\tfrac{\left( 1+\mu \right)
\sqrt{2\left( 1+\mu \right) }}{\sqrt{3+\mu }}\left( 1-\tfrac{\eta ^{2}}{%
a_{0}\nu \left( 1+\mu \right) }\right) \left( 1-\tfrac{\xi ^{2}}{a_{0}\nu
\left( 1+\mu \right) }\right) $ \\
20$\mu $ & $\sqrt{\tfrac{2\left( 1+\mu \right) ^{3}\left( 3+2\mu \right) }{%
3+\mu }} $ & $
\tfrac{\left( 1+\mu \right) \sqrt{2+\mu }}{\sqrt{3+\mu }} \left( 1-\tfrac{
2\xi ^{2}}{a_{0}\nu \left( 1+\mu \right) }+\tfrac{\xi ^{4}}{a_{0}^{2}\nu
^{2}\left( 2+\mu \right) \left( 1+\mu \right) }\right) $ \\
&$\times \left( 1-\tfrac{2r}{a_{0}\nu \left( 1+\mu \right) }+\tfrac{2r^{2}}{%
a_{0}^{2}\nu ^{2}\left( 1+\mu \right) \left( 3+2\mu \right) }\right)\bsep{5pt} $
& \\
 \hline\hline
\end{tabular}
\end{table}

Table~\ref{table6} illustrates directly the ratio of normalization constants and
polynomial factors for excited states in spherical and parabolic
coordinates, respectively.

The products of the spheroidal polynomials in equations~\eqref{eq68} for the excited
states with $n_{1}+n_{2}=1,2$, obtained from the recurrence relation of equation~\eqref{eq69}, have the same forms
\begin{gather}\label{eq72}
\left[ 1+\frac{A_{i}(u-1)}{2(1+\mu )}\right] \left[ 1+\frac{A_{i}(v-1)}{2(1+\mu )}\right]
\end{gather}
for the states $|n_{u}=0,n_{v}=1,\mu \rangle $ and $|n_{u}=1,n_{v}=0,\mu
\rangle $ where $A_{i}$ with $i=1,2$ are the ordered roots of the quadratic
equation
\begin{gather}\label{eq73}
A^{2}-\left( 2\mu +2-\frac{4f}{a_{0}\mu }\right) A-\frac{4f}{a_{0}\mu }%
\left( \mu +1\right) =0.
\end{gather}
Similarly, the states $|n_{u}=0,n_{v}=2,\mu \rangle $, $|n_{u}=1,n_{v}=1,\mu \rangle $ and $|n_{u}=2,n_{v}=0,\mu \rangle $ share the
products
\begin{gather}
\left[ 1+\frac{A_{i}(u-1)}{2(1+\mu )}-\frac{[A_{i}-2(1+\mu )](4f+A_{i}\nu
)(u-1)^{2}}{8(1+\mu )(2+\mu )\nu }\right]\nonumber\\
\qquad{}\times \left[ 1+\frac{A_{i}(v-1)}{2(1+\mu
)}-\frac{[A_{i}-2(1+\mu )](4f+A_{i}\nu )(v-1)^{2}}{8(1+\mu )(2+\mu )\nu }
\right],\label{eq74}
\end{gather}
where $A_{i}$ with $i=1,2,3$ are the ordered roots of the cubic equation
\begin{gather}
A^{3}+\left(-8-6\mu +\frac{12f}{a_{0}\nu }\right)A^{2}+\left[12+20\mu +8\mu ^{2}-\frac{%
16f(3\mu +4)}{a_{0}\mu }+\frac{12f^{2}}{(a_{0}\mu )^{2}}\right]A\nonumber\\
\qquad {}+\frac{16f(1+\mu
)(2\mu +3)}{a_{0}\mu }-\frac{64f^{2}(\mu +1)}{(a_{0}\mu )^{2}}=0.\label{eq75}
\end{gather}

The generalization for any f\/ixed value of $n_{u}+n_{v}$ involves the
tridiagonal matrix representation of equation~\eqref{eq69} of dimension $(n_{u}+n_{v}+1)\times (n_{u}+n_{v}+1)$. The corresponding secular
determinant generalizing equations~\eqref{eq73} and~\eqref{eq75} is an algebraic equation for
the eigenvalues~$A_{i}$ of degree $n_{u}+n_{v}+1$. The diagonalization of
the matrix yields the numerical values of~$A_{i}$'s and the coef\/f\/icients $c_{s}(A_{i})$ for the respective polynomials in equations~\eqref{eq68}.

\section{Raising and lowering operators}\label{section6}

On the basis of the equivalence of the hydrogen atom eigenfunctions in
parabolic coordinates, equations~\eqref{eq64},~\eqref{eq65}, and the harmonic oscillator
eigenfunctions in the circular radial coordinates, equation~\eqref{eq16}, they also share
the same kind of raising and lowering operators of equations~\eqref{eq29} and~\eqref{eq30} with
the correspondences $a=-n_{\xi },-n_{\eta }$, $b=\mu +1$, $z=\xi
^{2}/\nu a_{0}\xi$, $\eta ^{2}/\nu a_{0}$,
\begin{gather*}
\left( \mu +1+n_{\xi }-z+z\frac{d}{dz}\right) M\left( -n_{\xi },\mu
+1,z\right)  =\left( \mu +1+n_{\xi }\right) M\left( -\left( n_{\xi }+1\right)
,\mu +1,z\right),%\label{eq76}
\\
\left( -n_{\xi }+z\frac{d}{dz}\right) M\left( -n_{\xi },\mu +1,z\right)
 =-n_{\xi }M\left( -\left( n_{\xi }-1\right) ,\mu +1,z\right).%\label{eq77}
\end{gather*}

In spherical coordinates, both the hydrogen atom and the harmonic oscillator
share the same polar angle eigenfunctions of the square of the angular
momentum and $z$-parity, equation~\eqref{eq20},  as well as their raising and lowering
operators, in equations~\eqref{eq33}--\eqref{eq36}. For the hydrogen atom radial eigenfunctions of
equation~\eqref{eq58}, the corresponding operators also follow from equations~\eqref{eq29},~\eqref{eq30} using
the correspondences $a=-n_{r}$, $b=2\lambda +2$, $z=2r/\nu a_{0}$:
\begin{gather*}
\left( 2\lambda +2+n_{r}-z+z\frac{d}{dz}\right)\! M ( -n_{r},2\lambda
+2,z ) % \nonumber\\ \qquad{}
= ( 2\lambda +2+n_{r} ) M ( - ( n_{r}+1 )
 ) +2 ( 2\lambda +2,z ),%\label{eq78}
\\
\left( -n_{r}+z\frac{d}{dz}\right) M ( -n_{r},2\lambda +2,z )
 =-n_{r}M ( - ( n_{r}-1 ) ,2\lambda +2,z ).%\label{eq79}
\end{gather*}

The interested readers can easily check their connecting actions between the
successive entries in Table~\ref{table5}, including their extension.

{\sloppy Concerning the raising and lowering operators for the spheroidal polynomials
of equa\-tions~\eqref{eq68}, the inspection of their explicit forms for the lowest excited
states in equations~\eqref{eq72}--\eqref{eq75} lead us to their respective identif\/ications. In fact,
for each f\/ixed value of $n_{u}+n_{v}$, the $n_{u}+n_{v}+1$ degenerate
eigenfunctions share the same form dif\/fering in the associated eigenvalue~$A_{i}$. Consequently, the successive changes in the neighboring eigenvalues $A_{1}\rightarrow A_{2}\rightarrow A_{3}\rightarrow \cdots \rightarrow
A_{n_{u}+n_{v}}\rightarrow A_{n_{u}+n_{v}+1}$ translates into the raisings
for the spheroidal quantum number $n_{u}=0\rightarrow n_{u}=1\rightarrow
n_{u}=2\rightarrow \cdots \rightarrow n_{u}=n_{u_{\max}}-1\rightarrow
n_{u}=n_{u_{\max}}$ with the corresponding lowerings for the hyperboloidal
quantum number $n_{v_{\max}}\rightarrow n_{v_{\max}}-1\rightarrow
n_{v_{\max}}-2\rightarrow \cdots \rightarrow n_{v}=1\rightarrow n_{v}=0$, in such
way that $n_{u_{\max}}=n_{v_{\max}}=n_{u}+n_{v}$.

}

On the other hand, for the polynomial solutions in equations~\eqref{eq68}, their common
expansion coef\/f\/icients depend on the respective eigenvalue $c_{s}(A_{n_{u}}) $ where we replace the index $i=1,2,\dots, N_{\max }+1$
ordering the eigenvalues with the quantum number n$_{u}=0,1,2,\dots, N_{\max }$
counting the number of spheroidal nodes. Notice that both polynomials have
the same form and are of degree $n_{u}+n_{v},$ their dif\/ference being
associated with the distribution of their nodes in the spheroidal and
hyperboloidal degrees of freedom, in a complementary way. Their total
number is n$_{u}+n_{v}+1$ associated with states of the same energy. The
powers of $u-1$ or $v-1$ in which they are expanded appear in the same
number in equations~\eqref{eq68}. Both f\/inite sets of linearly independent sets of
power functions can be expressed as linear superpositions of the
corresponding sets $S_{n_{u}} ( u ) $ and $S_{n_{v}} ( v )$, via the inverse transformation based on the inverse matrix of the
coef\/f\/icient $c_{s} ( A_{n_{u}} )$. These relationships have their
counterparts for the familiar special functions, in polynomial and inf\/inite
series forms.

\section{Interbasis expansions of degenerate eigenfunctions\\ in spherical,
parabolic and prolate spheroidal coordinates}\label{section7}

Following the same order as in Section~\ref{section4}, the normalized eigenstates without
excitations in the non-common degrees of freedom are the same in the three
coordinates systems:
\begin{gather*}
\vert n_{r}=0, n_{\theta }=0,\mu \rangle ^{S}=\vert n_{\xi
}=0, n_{\eta }=0,\mu \rangle ^{P}=\vert n_{u}=0, n_v=0,\mu \rangle ^{PS}.
\end{gather*}

From the entries in Table~\ref{table6}, the connections between the normalized
eigenstates in spherical and parabolic coordinates become:
\begin{gather*}
\begin{pmatrix}
\left\vert 01\mu \rangle ^{S}\right. \vspace{3mm}\\
\left\vert 01\mu \rangle ^{S}\right.
\end{pmatrix}%
=
\begin{pmatrix}
\frac{1}{\sqrt{2}} & -\frac{1}{\sqrt{2}} \vspace{1mm}\\
\frac{1}{\sqrt{2}} & \frac{1}{\sqrt{2}}%
\end{pmatrix}
\begin{pmatrix}
\left\vert 01\mu \rangle ^{P}\right. \vspace{3mm}\\
\left\vert 01\mu \rangle ^{P}\right.%
\end{pmatrix} %\label{eq80}
\end{gather*}%
and%
\begin{gather*}%\label{eq81}
\begin{pmatrix}
|02\mu \rangle ^{S} \vspace{3mm}\\
|11\mu \rangle ^{S} \vspace{3mm}\\
|20\mu \rangle ^{S}%
\end{pmatrix}%
=
\begin{pmatrix}
\sqrt{\frac{1+\mu }{2\left( 3+2\mu \right) }} & -\sqrt{\frac{2+\mu }{3+2\mu }%
} & \sqrt{\frac{1+\mu }{2\left( 3+2\mu \right) }} \vspace{1mm}\\
\frac{1}{\sqrt{2}} & 0 & -\frac{1}{\sqrt{2}} \vspace{1mm}\\
\sqrt{\frac{2+\mu }{2\left( 3+2\mu \right) }} & \sqrt{\frac{1+\mu }{3+2\mu }}
& \sqrt{\frac{2+\mu }{2\left( 3+2\mu \right) }}%
\end{pmatrix}
\begin{pmatrix}
|02\mu \rangle ^{P} \vspace{3mm}\\
|11\mu \rangle ^{P} \vspace{3mm}\\
|20\mu \rangle ^{P}
\end{pmatrix},
\end{gather*}
where the orthonormality of the transformation matrices is obvious. These
are the counterparts of the interbasis expansions for the free hydrogen atom~\cite{19}.

Similarly, from the spherical entries in Table~\eqref{table6} and equations~\eqref{eq68} for the
prolate spheroidal polynomials, the transformation matrices between the
respective products of polynomials are
\begin{gather*}%\label{eq82}
\begin{pmatrix}
\frac{2\left( 1+\mu \right) }{f\left( A_{2}-A_{1}\right) }\left[ A_{2}-\frac{A_{2}^{2}}{2\left( 1+\mu \right) }\right] & -\frac{2\left( 1+\mu \right) }{f\left( A_{2}-A_{1}\right) }\left[ A_{1}-\frac{A_{1}^{2}}{2\left( 1+\mu
\right) }\right] \vspace{1mm}\\
\frac{A_{2}^{2}}{2\left( 1+\mu \right) \left( A_{2}-A_{1}\right) } & -\frac{A_{1}^{2}}{2\left( 1+\mu \right) \left( A_{2}-A_{1}\right) }
\end{pmatrix}
\end{gather*}
and
\begin{gather*}
\left[
\begin{array}{@{}cc}
-
\frac{8\left( 1+\mu \right) \left\{ 2\left( 1+\mu \right) \left[ 4+3\mu-\left( A_{2}+A_{3}\right) \right] +A_{2}A_{3}\right\} }
     {\left(A_{1}-A_{2}\right) \left( A_{1}-A_{3}\right) } &
\frac{8\left( 1+\mu \right) \left\{ 2\left( 1+\mu \right) \left[ 4+3\mu -\left( A_{3}+A_{1}\right) \right] +A_{3}A_{1}\right\} }
     {\left( A_{1}-A_{2}\right) \left(A_{1}-A_{3}\right) }
 \vspace{1mm}\\
-\frac{8f\left( 1+\mu \right) \left[ A_{2}-2\left( 1+\mu \right) \right] \left[ A_{3}-2\left( 1+\mu \right) \right] }
     {\left( A_{1}-A_{2}\right) \left( A_{1}-A_{3}\right) \left( A_{1}+2f\right) \left[ A_{1}-2\left( 1+\mu \right) \right] } &
\frac{8f\left( 1+\mu \right) \left[ A_{3}-2\left( 1+\mu \right) \right] \left[ A_{1}-2\left( 1+\mu \right) \right] }
     {\left(A_{1}-A_{2}\right) \left( A_{1}-A_{3}\right) \left( A_{1}+2f\right) \left[A_{1}-2\left( 1+\mu \right) \right] }
\vspace{1mm}\\
\frac{16f^{2}\left( 2+\mu \right) \left[ A_{2}A_{3}+4f\left( 1+\mu \right) \right] }
     {
         \left( A_{1}-A_{2}\right)
         \left( A_{1}-A_{3}\right)
         \left(A_{1}+2f\right)
         \left( 3+2\mu \right)
         \left[ A_{1}-2\left( 1+\mu \right) \right]
     } &
\frac{16f^{2}\left( 2+\mu \right) \left[ A_{1}A_{3}+4f\left(1+\mu \right) \right] }
     {\left( A_{1}-A_{2}\right) \left( A_{2}-A_{3}\right)\left( A_{2}+2f\right) \left( 3+2\mu \right) \left[ A_{2}-2\left( 1+\mu \right) \right] }
\end{array}\right.\nonumber\\
\left. \hspace{75mm}\begin{array}{c@{}}
\frac{8\left( 1+\mu \right) \left\{ 2\left( 1+\mu \right) \left[ 4+3\mu -\left( A_{3}+A_{1}\right) \right] +A_{3}A_{1}\right\} }
     {\left( A_{1}-A_{2}\right) \left( A_{1}-A_{3}\right) }\vspace{1mm}\\
 \frac{8f\left( 1+\mu \right) \left[A_{1}-2\left( 1+\mu \right) \right] \left[ A_{2}-2\left( 1+\mu \right) \right] }
     {\left( A_{1}-A_{2}\right) \left( A_{3}-A_{2}\right) \left(A_{3}+2f\right) \left[ A_{3}-2\left( 1+\mu \right) \right] }
 \vspace{1mm}\\
\frac{16f^{2}\left( 2+\mu \right) \left[A_{1}A_{2}+4f\left( 1+\mu \right) \right] }
     {\left( A_{1}-A_{3}\right) \left(A_{2}-A_{3}\right) \left( A_{3}+2f\right) \left( 3+2\mu \right) \left[ A_{3}-2\left( 1+\mu \right) \right] }
\end{array}
\right].%\label{eq83}
\end{gather*}
These interbasis expansions are the counterparts of those for the free
hydrogen atom~\cite{21, 20}.

\section{Discussion}\label{section8}

The dif\/ference between the free harmonic oscillator and hydrogen atom and
their conf\/ined by dihedral angle counterparts resides in the boundary
conditions of equations~\eqref{eq8},~\eqref{eq9} and~\eqref{eq50},~\eqref{eq51} in the respective coordinate systems.
The latter lead to the eigenfunctions and eigenvalues of the square of the
$z$-component of angular momentum of equation~\eqref{eq15}, where~$\mu $ replaces the
familiar integer magnetic quantum number $m$.

Nevertheless, both the free and the conf\/ined-by-dihedral-angle quantum
systems share their superintegrability and exact solvability. Admittedly,
the separability and integrability is restricted to the coordinate systems
sharing the dihedral angle coordinate: circular cylindrical and spherical
for the harmonic oscillator, and spherical, parabolic and prolate spheroidal
for the hydrogen atom. Sections~\ref{section3} and~\ref{section6} illustrated the identif\/ication and
construction of the raising and lowering operators for the eigenfunctions of~$L_{z}^{2}$, the transverse and axial contributions to the Hamiltonian of
the isotropic harmonic oscillator, its complete spherical radial
Hamiltonian, the square of the angular momentum and $z$-parity operators; and
their extensions and adaptations for the Hydrogen atom in the respective
coordinates. These operators have their free harmonic oscillator and
Coulomb potential counterparts in~\cite{22}. Concerning special functions, the
reader is invited to recognize the roles of the fractional Fourier basis,
equation~\eqref{eq15}, and the square of the angular momentum and $z$-parity eigenfunctions
equation~\eqref{eq20}. Additionally, Sections~\ref{section4} and~\ref{section7} illustrate the relationships
between the complete degenerate eigenfunctions of the conf\/ined quantum
systems in pairs of the alternative coordinate systems.

Some of the eigenfunctions and their raising and lowering operators may also
be adapted to other quantum or electromagnetic systems sharing the same
constants of motion and con\-f\/i\-ne\-ment situations. As specif\/ic examples the
free particle conf\/ined by dihedral angles, and nondif\/fracting vortex beams~\cite{23}, can be mentioned.

\pdfbookmark[1]{References}{ref}
\LastPageEnding

\end{document}